\documentclass[twocolumn,amsmath,amssymb,nofootinbib,preprintnumbers,prl]{revtex4}
\usepackage{color}
\usepackage{graphicx}% Include figure files
\usepackage{dcolumn}% Align table columns on decimal point
\usepackage{bm,psfrag}% bold math

\newcommand{\ben}{\begin{equation}}
\newcommand{\een}{\end{equation}}
\newcommand{\be}{\begin{equation}}
\newcommand{\ee}{\end{equation}}
\newcommand{\bea}{\begin{eqnarray}}
\newcommand{\eea}{\end{eqnarray}}
\newcommand{\ba}{\begin{eqnarray}}
\newcommand{\ea}{\end{eqnarray}}

\newcommand{\beq}{\begin{equation}}
\newcommand{\eeq}{\end{equation}}
\newcommand{\beqa}{\begin{eqnarray}}
\newcommand{\eeqa}{\end{eqnarray}}
\newcommand{\beqar}{\begin{eqnarray*}}
\newcommand{\eeqar}{\end{eqnarray*}}

\newcommand{\cO}{{\cal O}}

\def\t6 {T_\mt{D6}}

\newcommand{\tb}{\bar{M}}

\newcommand{\mt}[1]{\textrm{\tiny #1}}

\def\ta{\tilde{a}}
\def\tb{\tilde{b}}

\def\cale         {{\cal E}}

\def\vx{{\vec x}}

\def\ee           {{\rm e}}

\def\sqr#1#2{{\vcenter{\vbox{\hrule height.#2pt
 \hbox{\vrule width.#2pt height#1pt \kern#1pt
 \vrule width.#2pt}\hrule height.#2pt}}}}

\def\ee{\cale}

\def\aa1{\phi}
\def\cc1{\psi}

\def\ta{\tilde{a}}

\def\tE{\tilde{E}}

\def\ta{\tilde{a}}

\newcommand{\dt}{\delta t}

\usepackage[bookmarks=false]{hyperref} 
\hypersetup{pdfstartview=FitH,pdfhighlight=/O,colorlinks=false}

\begin{document}

\preprint{EFI-17-4,YITP-17-16,UK/17-02}

\title{Quantum Quench and Scaling of Entanglement Entropy}

\author{Pawe{\l} Caputa,$^{1}$ Sumit R. Das,$^{2}$ Masahiro
  Nozaki$^{3}$ and Akio Tomiya$^4$} 
\affiliation{$^1$\,Yukawa
  Institute for Theoretical Physics, Kyoto University, Kyoto 606-8502,
  JAPAN} 
\affiliation{$^2$\,Department of Physics and Astronomy,
  University of Kentucky, Lexington, KY 40506, USA}
\affiliation{$^3$\,Kadanoff Center for Theoretical Physics, University
  of Chicago, Chicago, IL 60637, USA} 
\affiliation{$^4$\, Key
  Laboratory of Quark \& Lepton Physics (MOE) and Institute of
  Particle Physics, Central China Normal University, Wuhan 430079,
  CHINA}

%\date{\today}
%%%%%%%%%%%%
\begin{abstract}
%%%%%%%%%%%%
Global quantum quench with a finite quench rate which crosses critical points is known to lead to universal scaling of correlation functions as functions of the quench rate. 
In this work, we explore scaling properties of the entanglement entropy of a subsystem in a harmonic chain during a mass quench which asymptotes to finite constant values at early and late times and for which the dynamics is exactly solvable. When the initial state is the ground state, we find that for large enough subsystem sizes the entanglement entropy becomes independent of size. This is consistent with Kibble-Zurek scaling for slow quenches, and with recently discussed "fast quench scaling" for quenches fast compared to physical scales, but slow compared to UV cutoff scales.

\end{abstract}

\maketitle

%\tableofcontents
%%%%%%%%%%%%%%%%%%%%%%%%
\noindent \textbf{1. Introduction:}
%%%%%%%%%%%%%%%%%%%%%%%%
The behavior of entanglement of a many-body system
that undergoes a quantum quench has been a subject of great interest in
recent times. When the quench is instantaneous (i.e. a sudden change of
the hamiltonian), several results are known. Perhaps the best known
result pertains to the entanglement entropy (EE) of a region of size
$l$ in a $1+1$ dimensional conformal field theory following a global
instantaneous quench, $S_{EE}(l)$. As shown in \cite{cc1}, $S_{EE}(l)$
grows linearly in time till $t \approx l/2$ and then saturates to a
constant value typical of a thermal state - a feature which has been
studied extensively in both field theory and in holography. Generalisations of this result to conserved charges and higher
dimensions have been discussed more recently
\cite{Caputa:2013eka,tsunami,mezei}. The emphasis of these studies is to probe the
     {\em time evolution} of the entanglement entropy.\\ 
     In physical situations, quantum quench has a finite rate, characterized by a
     time scale $\dt$, that can vary from very small to very large. When the quench involves a critical point,
     universal scaling behavior has been found for correlation
     functions at {\em early} times. The most famous scaling
     appears for a global quench which starts from a massive
     phase with an initial gap $m_g$, crosses a critical point (chosen
     to be e.g. at time $t=0$) and ends in another massive phase. For
     {\em slow quenches} (large $\delta t$), it has been conjectured that quantities obey
     Kibble-Zurek scaling \cite{kz}: evidence for this has been found
     in several solvable models and in numerical simulations
     \cite{kz2,dziarmaga}. Such scaling follows from two
     assumptions. First, it is assumed that as soon as the initial
     adiabatic evolution breaks down at some time $-t_{KZ}$ (the
     Kibble-Zurek time) the system becomes roughly diabatic. Secondly,
     one assumes that the only length scale in the critical region is
     the instantaneous correlation length $\xi_{KZ}$ at the time $t =-
     t_{KZ}$. This implies that, for example, one point functions
     scale as $\langle\cO (t)\rangle \sim \xi_{KZ}^{-\Delta}$, where $\Delta$
     denotes the conformal dimension of the operator $\cO$ at the
     critical point. An improved conjecture involves scaling
     functions. For example, one and two point correlation functions
     are expected to be of the form \cite{Zurek96,Cincio,Deng,DamskiZurek,DziarmagaRams,Kolodrubetz,Chandran} 
     \bea 
     \langle\cO
     (t)\rangle& \sim & \xi_{KZ}^{-\Delta}~F(t/t_{KZ}) \nonumber
     \\ \langle\cO(\vx,t)\cO(\vx^\prime,t^\prime)\rangle & \sim &
     \xi_{KZ}^{-2\Delta} F
     \left[\frac{|\vx-\vx^\prime|}{\xi_{KZ}},\frac{(t-t^\prime)}{t_{KZ}}\right]
\label{0-2a}
\eea
Some time ago, studies of slow quenches in AdS/CFT models have led to
some insight into the origin of such scaling without making these
assumptions \cite{holokz}.

For protocols in relativistic theories which asymptote to constant
values at early times, one finds a different scaling behavior in the
regime $\Lambda_{UV}^{-1} \ll \dt \ll m_{phys}^{-1}$, where
$\Lambda_{UV}$ is the UV cutoff scale, and $m_{phys}$ denotes any
physical mass scale in the problem. For example, 
\ben \langle\cO (t)\rangle \sim
\dt^{d-2\Delta}
\label{0-2}
\een
where $d$ is the space-time dimension. This "fast quench scaling"
behavior was first found in holographic studies \cite{blmv} and
subsequently shown to be a completely general result in any
relativistic quantum field theory \cite{dgm}. The result follows from
causality, and the fact that in this regime linear response becomes a
good approximation. Finally, in the limit of an instantaneous quench,
suitable quantities saturate as a function of the rate : for quench to
a critical theory a rich variety of universal results are known in
$1+1$ dimensions \cite{cc2}.

Much less is known about the behavior of entanglement and Renyi
entropies {\em as functions of the quench rate} - a key ingredient of universality. This has been
studied for the 1d Ising model (and generalizations) with a transverse
field which depends {\em linearly} on time, $g(t) = 1 -
\frac{t}{\tau_Q}$ \cite{moore}-\cite{Cincio,Francuz}. The system is prepared
in the instantaneous ground state at some initial time, crossing
criticality at $t=0$. The emphasis of \cite{moore} and \cite{Cincio,Francuz}
is on the slow regime, which means $\tau_Q \gg a$ where $a$ is the
lattice spacing, while \cite{canovi} also studies smaller values of
$\tau_Q$.  In particular, \cite{moore} and \cite{canovi} studied the
EE for half of a finite chain and found that the answer approaches
$S_{EE} \sim \frac{1}{12}\log \xi_{KZ}$ after sufficiently slow
quenches.  This is consistent with the standard assumptions which lead
to Kibble-Zurek scaling mentioned above.  According to these
assumptions, the system evolves adiabatically till $t = -t_{KZ}$ and
enters a phase of diabatic evolution soon afterwards. Thus the state
of the system at $t=0$ is not far from the ground state of the
instantaneous hamiltonian at $t = -t_{KZ}$.  Furthermore when $\tau_Q
\gg 1$ in lattice units, $\xi_{KZ}$ is large, and the instantaneous
state is close to criticality.  In such a state, the entanglement
entropy of a subregion of a large chain with $N_A$ boundary points
should obey an "area law" $\frac{c}{6} N_A \log(\xi_{KZ})$, where $c$
is the central charge. When the subsystem is half space $N_A=1$ and
for the Ising model the central charge is $c=1/2$.  Similarly,
\cite{Cincio,Francuz} studied the EE of a subsystem of finite size $l$ in an
infinite 1d Ising model, with a transverse field linear in time,
starting with the ground state at $t= -\infty$. The EE close to the
critical point for $l \gg \xi_{KZ}$ was found to saturate to $S_{EE} =
(\rm{constant}) +\frac{1}{6}\log (\kappa (t) \xi_{KZ})$. The factor
$\kappa (t) $ depends mildly on the time of measurement and $\kappa
(-t_{KZ}) \approx 1$. Once again, this result is roughly that of a
stationary system with correlation length $\xi_{KZ}$, as would be
expected from Kibble-Zurek considerations. The factor $\kappa(t)$ is a
correction to the extreme adiabatic-diabatic assumption. The paper
\cite{canovi} investigates an intermediate regime of fast quench (as
described above). While this paper investigates scaling of $S_{EE}$ as a function of quench rate in the slow regime, there is no similar analysis in the fast regime.

In this letter, we study entanglement entropy for a simple system: an
infinite harmonic chain (i.e. a 1+1 dimensional bosonic theory on a
lattice) with a time dependent mass term which
asymptotes to constant {\em finite} values at early and late times. We
choose a mass function for which the quantum dynamics can be solved
exactly. The use of such a protocol allows us to explore the whole
range of quench rates, where the speed of quench is measured in units
of the initial gap rather than the lattice scale.  We compute the
entanglement entropy for a subsystem of size $l$ (in lattice units) in
the middle of the quench and find that it scales in interesting ways
as we change the quench rate. The dimensionless quantity which
measures the quench timescale is $\Gamma_Q = m_0 \dt$ where $m_0$ is the
initial gap.  

%%%%%%%%%%%%%%%%%%%%%%%%
\noindent \textbf{2. Our setup and quench protocols:}
%%%%%%%%%%%%%%%%%%%%%%%%
The hamiltonian of the harmonic chain is given by 
\ben H = \frac{1}{2}
\sum_{n=-\infty}^{\infty} \left[ P_n^2 + (X_{n+1} - X_n)^2 + m^2 (t)
  X_n^2 \right]
\label{1-1}
\een
where $(X_n,P_n)$ are the usual canonically conjugate scalar field
variables on an one dimensional lattice whose sites are labelled by
the integer $n$. The mass term $m(t)$ is time dependent. All
quantities are in lattice units. In terms of momentum variables
$X_k,P_k$ 
\ben X_n (t) = \int_{-\pi}^{\pi} \frac{dk}{2\pi} X_k
(t)~e^{ikn}~~~~~~ P_n (t) = \int_{-\pi}^{\pi} \frac{dk}{2\pi} P_k (t)
~e^{ikn}
\label{1-2}
\een
the equation of motion is given by
\ben
\frac{d^2 X_k}{dt^2} + [4 \sin^2 (k/2) + m^2(t) ] X_k = 0
\label{1-3}
\een
We are interested in functions $m(t)$ which asymptote to constant
values $m_0$ at $t \rightarrow \pm \infty$, and pass through zero at
$t=0$.  Let $f_k (t)$ be a solution of (\ref{1-3}) which asymptotes to
a purely positive frequency solution $\sim e^{-i\omega_0
  t}/\sqrt{2\omega_0}$ at $t \rightarrow -\infty$, where
\ben
\omega_0^2 = 4 \sin^2 (k/2) + m_0^2.
\een
 A mode decomposition 
\ben
X_k (t) = f_k (t) a_k + f^\star_k (t) a^\dagger_{-k}
\label{1-4}
\een
with $ [ a_k , a^\dagger_{-k^\prime} ] = 2\pi \delta (k - k^\prime)$
can be then used to define the "in" vacuum by $a_k |0\rangle = 0$ for
all $k$. The solutions $f_k (t)$ are chosen to satisfy the Wronskian
condition $f_k ({\dot{f_k}})^\star - ({\dot{f_k}}) f_k^\star = i$. The
state $|0\rangle$ then denotes the Heisenberg picture ground state of
the initial Hamiltonian.  The normalized wavefunctional for the "in" vacuum state is given by
\ben
\Psi_0 (X_k,t) =\prod_k 
\frac{1}{[\sqrt{2\pi}f^\star_k(t)]^{1/2}}~{\rm{exp}}\left[ \frac{1}{2} \left(\frac{{\dot{f_k(t)}}}{f_k(t)}\right)^\star X_k X_{-k} \right]
\label{1-5}
\een

We will choose a quench protocol for a mass function for which the mode functions $f_k(t)$ can be solved exactly. The particular mass function we use is
\ben
m^2(t) = m_0^2 \tanh^2 (t/\dt)
\label{1-6}
\een
The corresponding mode functions are given by
\begin{eqnarray}
&&\!\!\!\!
f_k  =\frac{1}{\sqrt{2\omega_0}}\frac{(2)^{i\omega_0 \dt}\cosh^{2\alpha} (t/\dt)}{E^\prime_{1/2}\tE_{3/2}-E_{1/2}\tE^\prime_{3/2}} \times \nonumber \\
&&\!\!\!\!
[ \tE^\prime_{3/2}~ _2F_1 (\ta,\tb,\frac{1}{2};-\sinh^2(t/\dt))  \nonumber \\
&&\!\!\!\!
+ E_{1/2}^\prime \sinh(t/\dt)~_2F_1 ( \ta+\frac{1}{2}, \tb+ \frac{1}{2},\frac{3}{2};-\sinh^2(t/\dt)) ]
\nonumber
\end{eqnarray}
where we have defined 
\bea
\omega_0^2 (k) & = & 4 \sin^2 (k/2) + m_0^2 \nonumber \\
\alpha & = & \frac{1}{4}[1+\sqrt{1-4m_0^2\dt^2}] \nonumber \\
\ta & = & \frac{1}{4}[1+\sqrt{1-4m_0^2\dt^2}]+\frac{i}{2}\dt\omega_0 \nonumber \\
\tb & = & \frac{1}{4}[1+\sqrt{1-4m_0^2\dt^2}]-\frac{i}{2}\dt\omega_0
\nonumber \\
E_{1/2} & = & \frac{\Gamma(1/2)\Gamma(\tb- \ta)}{\Gamma(\tb)\Gamma(1/2-\ta)} \nonumber \\
\tE_{3/2} & = & \frac{\Gamma(3/2)\Gamma(\tb - \ta)}{\Gamma(\tb+1/2)\Gamma(1-\ta)} \nonumber \\
E^\prime_c & = & E_c ( \ta \leftrightarrow \tb).
\label{1-8}
\eea

%%%%%%%%%%%%%%%%%%%%%%%%
\noindent \textbf{3. Entanglement Entropy:} 
%%%%%%%%%%%%%%%%%%%%%%%%
Our aim is to calculate the entanglement entropy for a subregion of the infinite chain consisting of $l$ lattice points. This is most conveniently calculated by considering the $2l \times 2l$ matrix of correlators, and the symplectic matrix \cite{correntangle,mezei}
\[
C =
\begin{bmatrix}
X_{mn}(t) & D_{mn}(t) \\
D_{mn}(t) & P_{mn}(t) 
\end{bmatrix}
~~~~
J = \begin{bmatrix}
0 & I_{l\times l} \\
-I_{l \times l} & 0
\end{bmatrix}
\]
where $(m,n)$ denote lattice sites inside the subsystem and 
\bea
X_{mn}(t) & = & \langle0|X_m(t) X_n(t)|0\rangle,~~
P_{mn}(t)  =  \langle0|P_m(t)P_n(t)|0\rangle, \nonumber \\
D_{mn}(t) & = &\frac{1}{2}\langle0|\{X_m(t),P_n(t)\}|0\rangle
\label{correl}
\eea
The eigenvalues of the matrix $iJC$ then occur in pairs $\pm \gamma_n(t) ~~n=1,\cdots l$ where $\gamma_n (t) > 0$. The entanglement entropy $S$ is given by
\ben
S  = \sum_{n=1}^l \{ [\gamma_n(t) + \frac{1}{2}]\log [\gamma_n(t)+\frac{1}{2}] - [\gamma_n(t) - \frac{1}{2}]\log [\gamma_n(t)-\frac{1}{2}] \}
\label{3-2}
\een

In terms of the mode functions (\ref{1-8}), the correlators (\ref{correl}) are given by the expressions
\bea
X_{mn}(t) & = & \int_{-\pi}^\pi \frac{dk}{2\pi} |f_k(t)|^2 \cos (k|m-n|) \nonumber \\
P_{mn}(t) & = & \int_{-\pi}^\pi \frac{dk}{2\pi} |{\dot{f}}_k(t)|^2 \cos (k|m-n|) \nonumber \\
D_{mn}(t) & = & \int_{-\pi}^\pi \frac{dk}{2\pi} {\rm{Re}}\left(f^*_k(t) {\dot{f}}_k(t) \right) \cos (k|m-n|)\nonumber\\
\eea

The correlator $D_{mn}(t)$ is particularly important in the calculations which follow. This is because this quantity is exactly zero when the mass is time independent. For the same reason, this quantity vanishes in the leading order of the adiabatic expansion (i.e. when observables are replaced by their {\em static} answers with the instantaneous value of the mass), i.e. $ f^{adia}_k = \frac{1}{\sqrt{2\omega_k(t)}}
e^{-i\omega_k(t)t} $ with $ \omega_k(t)^2 = 4 \sin^2 (k/2) + m^2(t) $.

The computation of correlation functions and the entanglement entropy then involves integrals over momenta $k$, which are computed numerically. As is well known, numerical integrals with oscillating functions in the integrand are tricky. Numerical methods reduce this to a summation of
an alternating series, leading to slow convergence. 
To confirm our numerical integration,  we check the convergence of the integration with varying precision. We perform all calculations with interval sizes $dk = 0.00001, 0.000001$.

%%%%%%%%%%%%%%%%%%%%%%%%
\noindent \textbf{4. Regimes of the quench rate:} 
%%%%%%%%%%%%%%%%%%%%%%%%
Before we examine these entropies, it is necessary to understand the various regimes of the dimensionless parameter $\Gamma_Q \equiv m_0 \dt$. This can be done by studying the scaling behavior of correlation functions.  Since we are interested in the slow as well as the fast regime, we will stay close to the continuum limit where the dimensionless mass $m_0 a$ is small. In the following we express everything in lattice units. Then, Kibble Zurek scaling is expected to hold for these quantities for $\Gamma_Q \gg 1$, while we get the regime of fast quench when $\Gamma_Q \ll 1$, but $\dt > 1$.

\begin{figure}[h]
\begin{center}
 \includegraphics[scale=0.4]{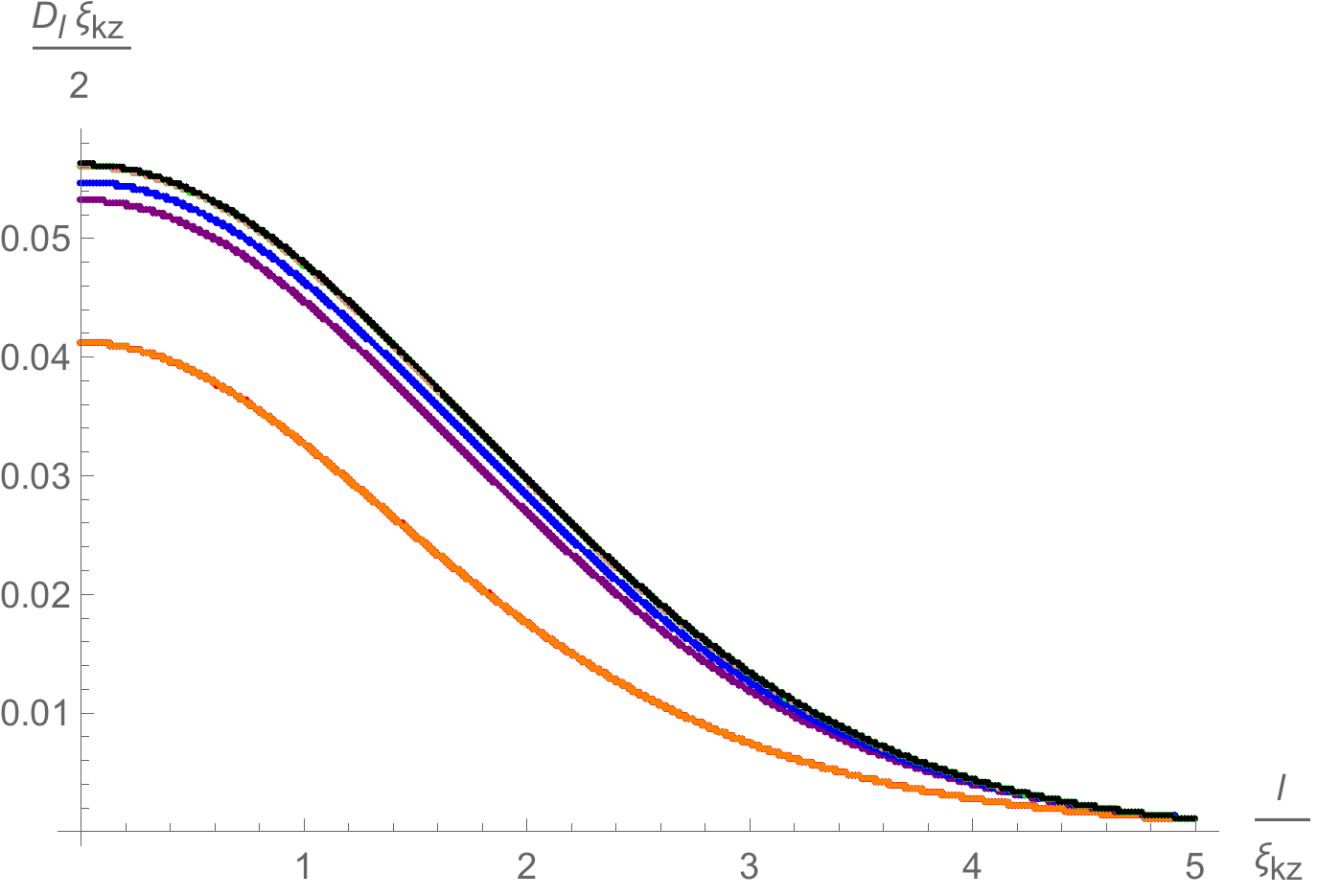}
\end{center}
\caption{(Color Online) Rescaled correlator $\frac{1}{2}\xi_{KZ}D_{mn}(t=0)$ as a function of $|m-n|$. The various values of $\Gamma_Q, \xi_{KZ}$ are in different colors:
On top of each other we see Red: ($\Gamma_Q = 1, \xi_{kz} =400$) and Orange: ($\Gamma_Q = 1, \xi_{kz} =500$), then Purple: ($\Gamma_Q =5, \xi_{kz} =400$) and Blue: ($\Gamma_Q =10, \xi_{kz} =400$) and again on top of each other: Green: ($\Gamma_Q=100, \xi_{kz} =400$), Pink: ($\Gamma_Q=100, \xi_{kz} =500$) and Black: ($\Gamma_Q=500, \xi_{kz} =500$).}
\label{corr}
\end{figure}

For slow quench, the usual Landau criterion leads to a Kibble Zurek time and the instantaneous correlation length at this time to be $
t_{KZ} = \xi_{KZ} = \sqrt{\dt / m_0}$,
and we expect e.g. the following leading answers at $t=0$ : $
X_{mn}(0)  \sim  \xi_{KZ}^{0} F_X(|m-n|/\xi_{KZ}), 
P_{mn}(0)  \sim  \xi_{KZ}^{-2} F_P(|m-n|/\xi_{KZ}),
D_{mn} (0)  \sim  \xi_{KZ}^{-1}F_D(|m-n|/\xi_{KZ})$,
where $F_X,F_P,F_D$ are smooth functions.
We indeed find this behavior for $\Gamma_Q \geq 100$. For example, Figure (\ref{corr}) shows $ \xi_{KZ} D_{mn} (0)$ as a function of $|m-n|$ for various values of $\Gamma_Q$. Clearly for large enough $\Gamma_Q$ the points fall on top of each other, showing that the above scaling behavior holds. The results for $X_{mn}(0)$ and $\xi_{KZ}^2 P_{mn}(0)$ are also consistent with the above expectations.

In the fast quench regime $\Gamma_Q \ll 1$ the analysis of \cite{dgm} leads to the following scaling relations for coincident correlators
\ben
X_{nn} \sim ({\rm{const}}) + (\dt)^2~~~
P_{nn} \sim ({\rm{const}}) + (\dt)^0~~~
D_{nn} \sim \dt
\label{2-3}
\een
We found this behavior for small $\Gamma_Q < 0.1 $. For correlation functions $X_{mn},P_{mn},D_{mn}$, with the constant mass values subtracted, we expect these scalings to hold for large $ |m-n| < \dt < m_0^{-1}$, while for $\dt < |m-n| < m_0^{-1}$ they should become independent of $\dt$ \cite{dgm}. We indeed see this behavior for sufficiently small $\Gamma_Q$. This is most clearly seen for $D_{mn}$ since the constant mass value of this quantity is exactly zero.

\begin{figure}[h]
\begin{center}
\includegraphics[scale=0.4]{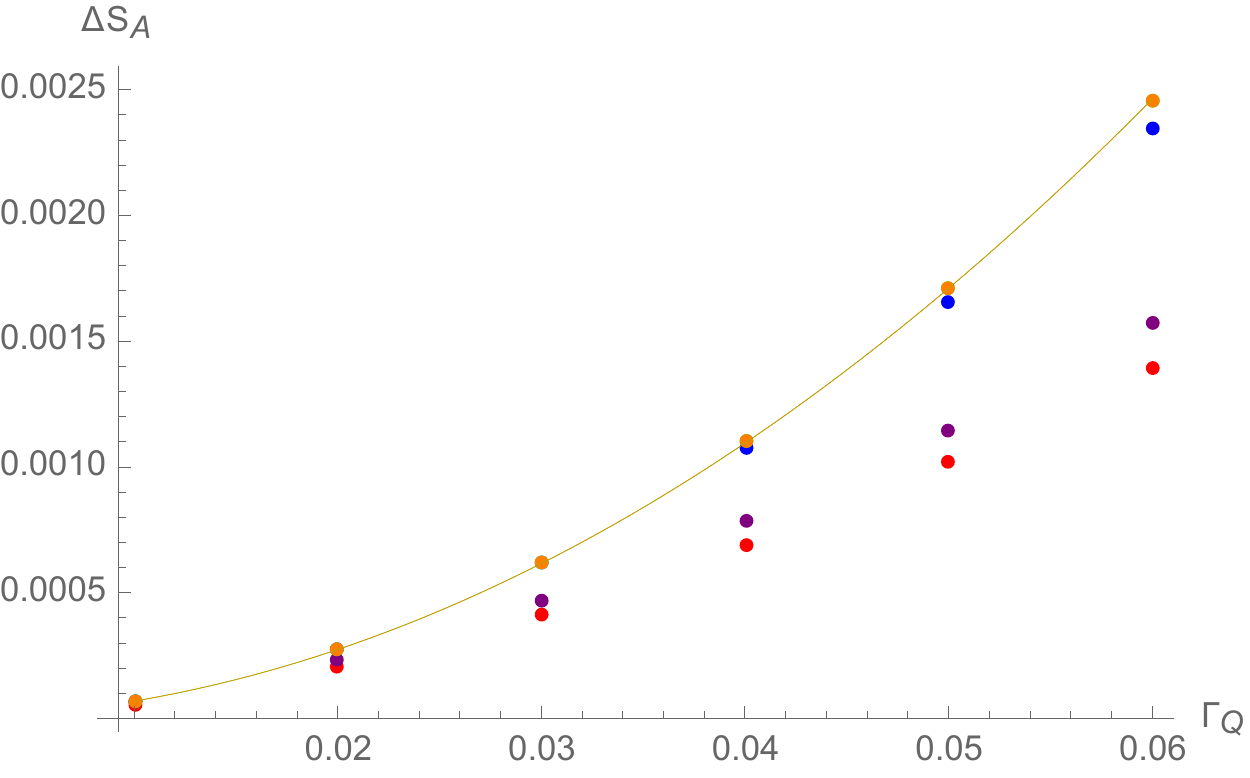}
\end{center}
\caption{(Color Online) $\Delta S_A$ at $t=0$ as a function of $\Gamma_Q$ for fast quench. All the data are for $\xi_i = 600$. Different colors correspond to different values of $l$. Red, Purple, Blue, Green and Orange have $l = 5,10,100,1000,2000$ respectively. The data for $l=1000,2000$ are almost on top of each other.}
\label{fast1}
\end{figure}

%%%%%%%%%%%%%%%%%%%%%%%%
\noindent \textbf{5. Scaling of entropies:} 
%%%%%%%%%%%%%%%%%%%%%%%%
Now that we have identified the slow and fast quench regimes, we go ahead and explore possible scaling properties of the entanglement entropy $S_{EE}$. This quantity is studied in detail at $t=0$.
(which is the middle of the quench) for both fast and slow quenches.\\
First, we consider the fast quench regime, $\Gamma_Q \ll 1$. We have computed the difference of the entanglement entropy at $t=0$ from its value at $t = -\infty$, 
\ben
\Delta S_A = S(t=0) - S(t=-\infty)
\een
for various subsystem sizes $l$ as a function of $\Gamma_Q$. Figure (\ref{fast1}) shows the results for $\xi_i=600$ as a function of $\Gamma_Q$. When $l$ is smaller than the initial correlation length $\xi_i= m_0^{-1}$ this quantity depends on $l$ for a given $\Gamma_Q$. However for $l$ sufficiently large compared to $\xi_i$ we find that the $l$-dependence saturates.

Namely, the data for large $l/\xi_i$ fits to the following result for small $\Gamma_Q$
\ben
\Delta S = c~\Gamma_Q^2, 
\label{r-1}
\een
where the constant $c$ is independent of $l$ and $\xi_i$ and approximately equal to $0.67$. 

\begin{figure}[h!]
\includegraphics[scale=0.4]{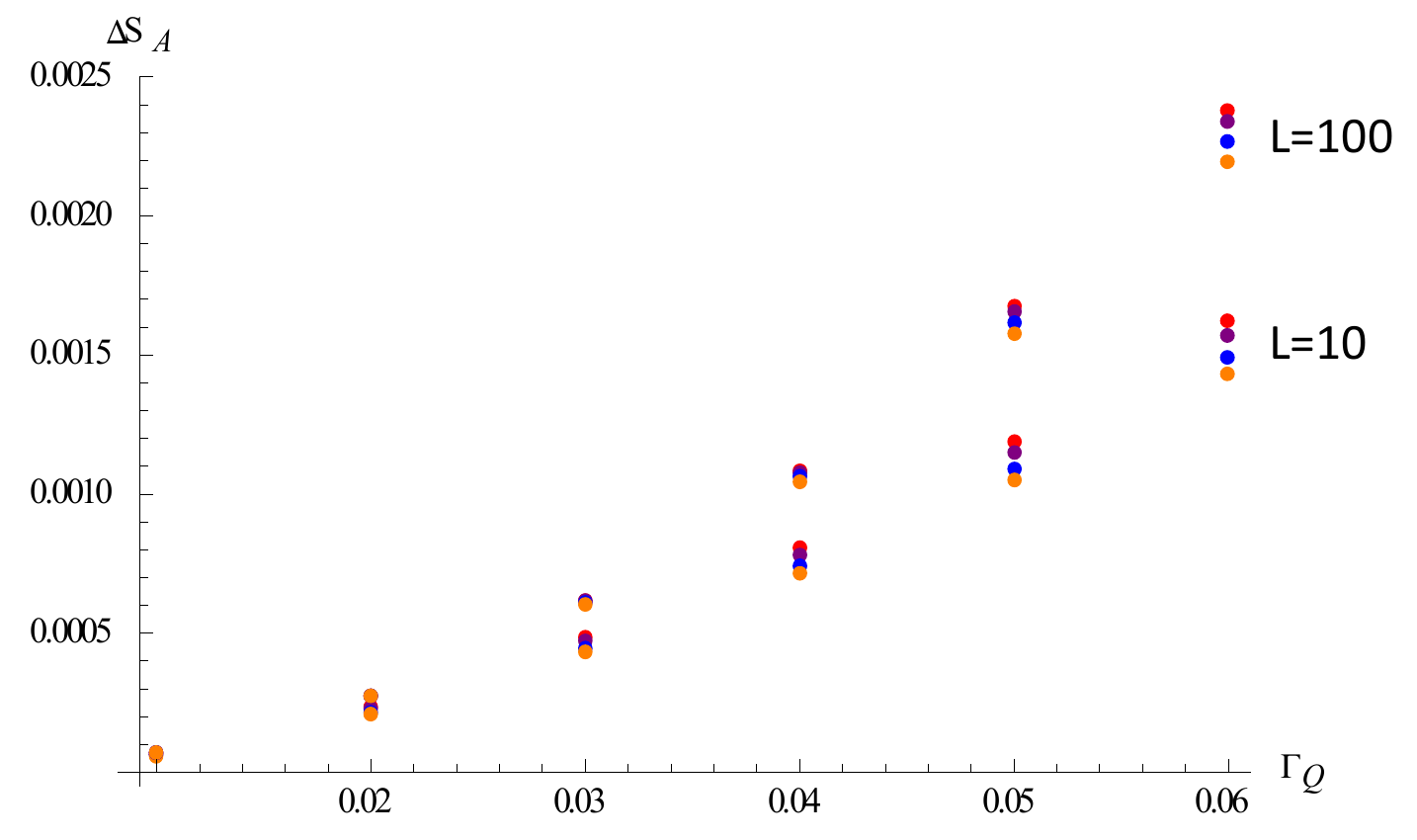}
\caption{(Color Online) $\Delta S_A$ at $t=0$ as a function of $\Gamma_Q$ for fast quench. The data are for $l=10$ and $l = 100$. Different colors correspond to different values of $\xi_i$. Red, Purple, Blue, and Orange have $\xi_i = 500, 600, 800, 1000$ respectively.}
\label{fast2}
\end{figure}

\begin{figure}[h!]
\begin{center}
\includegraphics[scale=0.4]{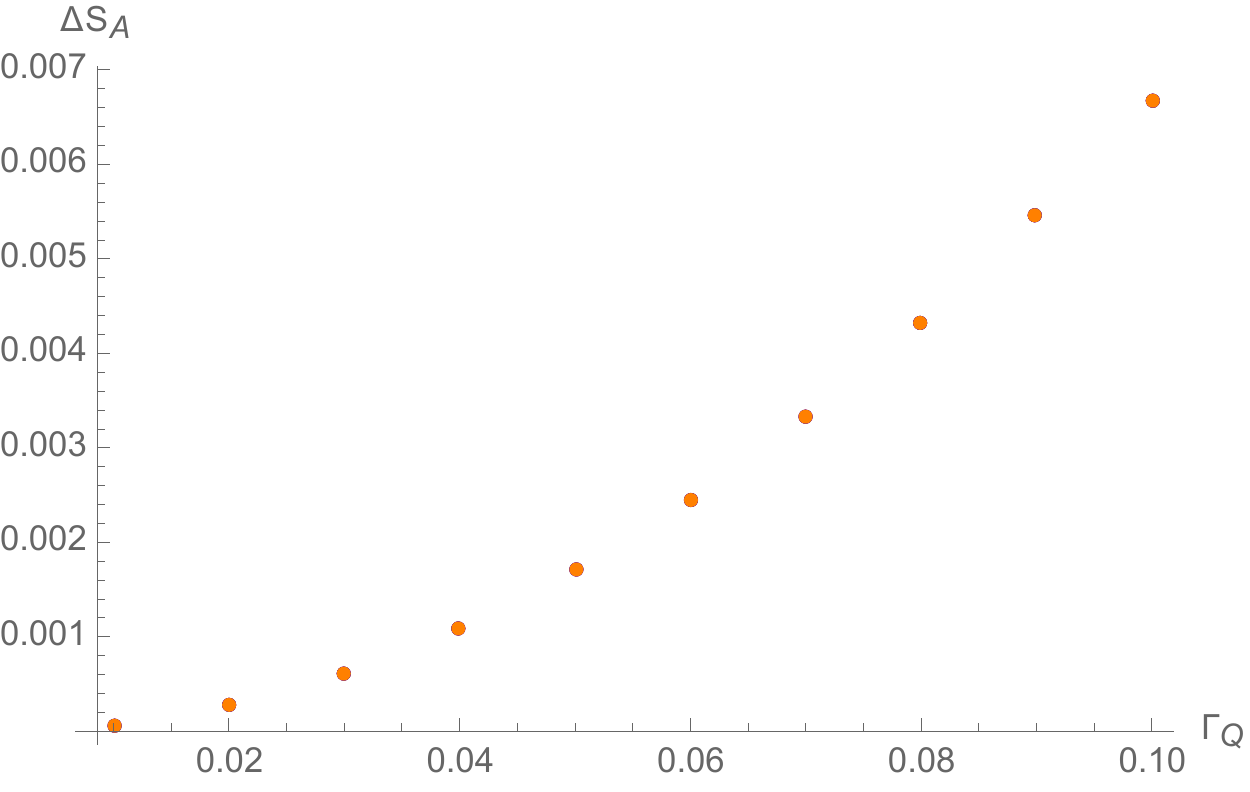}
\end{center}
\caption{(Color Online) $\Delta S_A$ at $t=0$ as a function of $\Gamma_Q$ for fast quench. All the data are for $l = 2000$. We plotted this for different values of $\xi_i$. Red, Purple, Blue, and Orange have $\xi_i = 500, 600, 800, 1000$ respectively. However the data are basically on top of each other. }
\label{fast3}
\end{figure}

In fact, for small $l/\xi_i$, $\Delta S_A$ does depend on $\xi_i$ for a given $\Gamma_Q$ and $l$, as shown in Figures (\ref{fast2}) and (\ref{fast3}). These are plots of $\Delta S_A$ as a function of $\Gamma_Q$ for a given $l$, and different values of $\xi_i$. Figure (\ref{fast2}) has $l=100$ with $\frac{l}{\xi_i}$ ranging from $0.1$ to $0.2$ : the result has a clear dependence on $\xi_i$. Figure (\ref{fast3}) has $l=2000$ with $\frac{l}{\xi_i}$ ranging from $2.0$ to $4.0$ - the data for various values of $\xi_i$ are right on top of each other.

This result (\ref{r-1}) is in agreement with expectations based on \cite{dgm}. In these papers it was argued that the fast quench scaling for one point functions follow from the fact that linear response becomes a good approximation in this regime.  Even though the scaling behavior of the two point function is complicated (as mentioned above), one would expect that the $\dt$ dependence of the matrix elements of $C$ would still be governed by perturbation theory. To lowest order this should be therefore proportional to $m_0^2$. In the regime of quench rates we are considering, the only other scale is $\dt$. Since the entanglement entropy is dimensionless, we expect that the leading answer is proportional to $m_0^2\dt^2$ which is $\Gamma_Q^2$.
The fact that the result becomes independent of $l$ is also expected since the measurement is made at $t=0$. 

%This is too early for excitations to spread over the system which could lead to an extensive term. Preliminary data for late times \cite{nozaki} indeed show that an extensive term begins to emerge.

It would be interesting to gain further insight by using the perturbation theory for entanglement entropy (as e.g. developed in \cite{smolkin}).

\begin{figure}[h!]
\begin{center}
\includegraphics[scale=0.4]{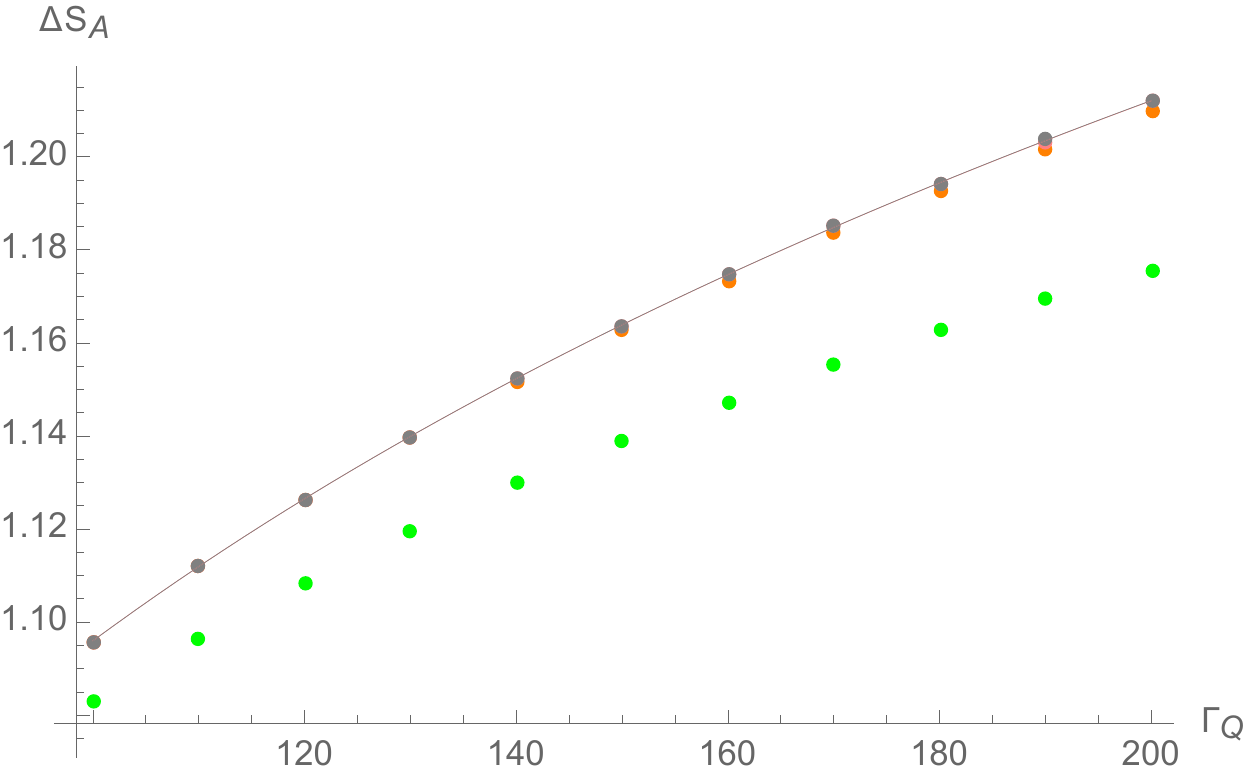}
\end{center}
\caption{(Color Online) $\Delta S_A$ at $t=0$ as a function of $\Gamma_Q$ for slow quench. The data are for $\xi_i = 40$ and $\xi_i=50$ for different values of $l$. However the points for these two different $\xi_i$ are basically on top of each other. Green, orange, pink and grey correspond to $l= 1000, 2000, 3000, 3500$. The $l=3000$ and $l=3500$ values are too close to be distingushable.}
\label{slow1}
\end{figure}

Let us now consider quench rates slow enough to ensure that the correlation functions discussed above display standard Kibble Zurek scaling. Figure (\ref{slow1}) shows $S_{EE}$ as a function of $\Gamma_Q$ for a given $m_0$ various values of subsystem size $l$.

For large enough $l$ we therefore find that $\Delta S_A$ is independent of $l$ and fits the relation
\ben
\Delta S_A = ({\rm constant}) + \frac{1}{6} \log (\Gamma_Q)
\label{4-1}
\een
The constant in (\ref{4-1}) is roughly $0.3$. This result is consistent with the expectations from Kibble Zurek scenario. More precisely, the state of the system at $t=0$ is fairly close to the instantaneous ground state at $t = -t_{KZ}$, i.e. the ground state of the system with a fixed (time independent) correlation length $\xi_{KZ}$. For such a system the entanglement entropy should behave as $\frac{1}{3} \log (\xi_{KZ})$, independent of $l$ for $l > \xi_{KZ}$. Since $\xi_{KZ}=\xi_i \sqrt{\Gamma_Q}$, this is $\frac{1}{3} \log (\xi_i) + \frac{1}{6} \log (\Gamma_Q)$.  On the other hand, $S_{EE} (t=-\infty)$ is ground state entanglement entropy of the system at a fixed correlation length $\xi_i$ and therefore behaves as $\frac{1}{3} \log (\xi_i)$ for $l \gg \xi_i$. Thus the KZ scenario predicts that when $\Delta S_A$ is expressed as a function of $\Gamma_Q$, the dependence on $\xi_i$ should cancel - which is exactly as we find.

In conclusion, we have calculated the entanglement entropy of a subsystem in a harmonic chain in the presence of an exactly solvable mass quench and examined the dependence of this quantity at the middle of the quench on the quench rate. Our nonlinear quench protocol asymptotes to constant values at initial and final times : in this respect it differs significantly from the kind of protocols most commonly studied in the literature, e.g. couplings which behave linearly in time.
For quench rates which are fast compared to the initial mass (but slow compared to the lattice scale) our results show, for the first time, that the fast quench scaling established for correlation functions in \cite{dgm} extend to non-local quantities like entanglement entropy. For  quench rates slow compared to the initial mass, our results are consistent with the predictions of Kibble-Zurek scenario, and with earlier results in the 1d Ising model.

\noindent \textbf{Note} In an earlier version of this paper, we had reported that the entanglement entropies have a piece which is extensive in the subsystem size. This result is erroneous and stemmed from a mistake in our numerical code. This revised version corrects the error.

%%%%%%%%%%%%%%%%%%%%%%%%
\noindent \textbf{Acknowledgements.} 
%%%%%%%%%%%%%%%%%%%%%%%%
We thank Shinsei Ryu, Marek Rams, Krishnendu Sengupta and Tadashi Takayanagi for discussions and comments on the manuscript. The work of PC is supported by the Simons Foundation through the "It from Qubit" collaboration. The work of S.R.D. is supported by a National Science Foundation grant NSF-PHY-1521045. The work of A.T. was supported in part by NSFC under grant number 11535012. A part of numerical computation in this work was carried out at the Yukawa Institute Computer Facility.

\end{document}